\documentclass[reprint, twocolumn, superscriptaddress,prb]{revtex4-2}
\usepackage{bm, amsmath, amsfonts, amssymb, braket}
\usepackage{subfigure}
\usepackage{color}
\usepackage{graphicx}
\usepackage{slashed}
\usepackage{etoolbox}
\apptocmd{\thebibliography}{\raggedright}{}{}

\usepackage{dcolumn} 

\usepackage{rotating} 

\usepackage{ulem}
\usepackage{xcolor}
\DeclareRobustCommand{\erase}{\bgroup\markoverwith{\textcolor{red}{\rule[.5ex]{2pt}{0.4pt}}}\ULon}

\begin{document}
\title{Topological two-body interaction obstructing trivial ground states: an indicator of fractional Chern insulators}
\author{Nobuyuki Okuma}
\email{okuma@hosi.phys.s.u-tokyo.ac.jp}
\affiliation{%
  Graduate School of Engineering, Kyushu Institute of Technology, Kitakyushu 804-8550, Japan
}%

\author{Tomonari Mizoguchi}
\affiliation{Department of Physics, University of Tsukuba, Tsukuba, Ibaraki 305-8571, Japan}

\date{\today}
\begin{abstract}
The search for candidate materials for fractional Chern insulators (FCIs) has mainly focused on the topological and geometrical structures of single-particle Chern bands. However, there are inherent limitations in approaches that neglect interaction effects, highlighting the need for complementary methods.
In this work, we discuss how the Chern number defined for the effective interaction projected onto a Chern band is related to the stabilization of FCIs. Specifically, by formulating both the effective interaction and the two-particle problem using a common matrix, we establish a connection between the two-particle band structure and the effective interaction. This formulation allows us to characterize the effective interaction through the topology of the two-particle band.
To investigate the relationship between topological effective interactions and FCIs, we perform numerical calculations primarily based on exact diagonalization.
We find a notable correlation between the fact that the dominant two-particle bands carry a unit Chern number and the realization of a robust FCI at the filling fraction $\nu = 1/3$.
This result is consistent with the presumed correspondence between pseudopotentials in the fractional quantum Hall effect and the two-particle band structure.
From another perspective, our findings suggest that the topology inherent in the interaction itself can obstruct trivial ground states.
We also discuss this in the context of scattering channels.
Extending such topological two-body interactions could pave the way for realizing exotic states beyond FCIs.

\end{abstract}
\maketitle

\section{Introduction}
The interplay between topology and interactions has recently emerged as one of the most actively studied areas in condensed matter physics. A pioneering example in this field is the fractional quantum Hall effect (FQHE) \cite{FQHE-exp-82,Laughlin-Wavefunction-83,yoshioka-textbook-02,halperin-fractional-textbook-20,fradkin2013field}. Under a strong magnetic field, Landau levels possess a topological electronic structure \cite{thouless1982quantized} and, at the same time, can be regarded as a kind of flat-band system due to their dispersionless nature. This allows each Landau level to serve as a strongly correlated topological system. The FQHE is realized at fractional fillings of such systems, where the ground states exhibit a type of topological order \cite{wen1990topological} that cannot be understood within a single-particle picture. In particular, the analytical properties of Landau levels have enabled the construction of useful variational wave functions, such as the Laughlin wave function \cite{Laughlin-Wavefunction-83}, which have greatly contributed to the understanding of FQHE physics.

More recently, there has been growing interest in fractional Chern insulators (FCIs) \cite{Regnault-Bernevig-11, Bergholtz-Liu-13, Parameswaran-13, Liu-Bergholtz-review-22}, which aim to realize the physics of the FQHE in systems without magnetic fields. 
Even in the absence of a magnetic field, Chern insulators can possess Chern bands that are topologically equivalent to Landau levels when the absolute value of the Chern number is one \cite{Haldane-88}. Therefore, it is natural to expect that fractional fillings of such Chern bands could give rise to states analogous to those of the FQHE. However, a clear limitation of this naive argument is that Chern bands do not necessarily have flat dispersion. Moreover, although the topological equivalence between Chern bands and Landau levels exists, it does not imply an equivalence in their one-particle electronic structures. As a result, the analytic properties that played a crucial role in the understanding of the FQHE is generally absent in Chern bands.

For these reasons, the search for candidate materials for FCIs has primarily focused on identifying ideal Chern bands that mimic the analytic properties of Landau levels \cite{parameswaran2012fractional,Roy-geometry-14,Jackson-Moller-Roy-15,Claassen-Lee-Thomale-Qi-Devereaux-15, Lee-Claassen-Thomale-17}. Ideal Chern bands can be formulated in terms of the quantum geometric tensor, and more recently, they have been described in a unified manner within the framework of vortexability \cite{ledwith2023vortexability,fujimoto2025higher}, which offers a more intuitive connection to the Landau levels. In the case of twisted two-dimensional materials, where the experimental realization of FCIs has been reported \cite{li2021spontaneous,cai2023signatures,zeng2023thermodynamic,park2023observation,xu2023observation,lu2023fractional}, it has also been argued that band structures exhibiting vortexability are realized \cite{ledwith2023vortexability}.

On the other hand, discussing the physics of strongly correlated materials based solely on single-particle band structures is insufficient, and alternative search approaches have also been explored. The simplest attempt to incorporate interactions is to examine the two-particle spectrum. Drawing on an analogy with Haldane's pseudopotentials \cite{Haldane-Hierarchy-83}, which have traditionally been used in studies of the FQHE, the dispersion structure formed by two particles on a Chern band has been investigated \cite{Lauchli-Liu-Bergholtz-Moessner-13,Liu-Bergholtz-Kapit-13}.
In our previous work \cite{okuma2023relationship}, we hypothesized that a certain type of Chern number defined in the two-particle band structure is related to the stabilization of bosonic FCIs in models with onsite repulsive interactions. In this paper, to make the hypothesis more convincing and broadly applicable, we present a more refined theoretical framework.
This new formalism allows us to treat both bosonic and fermionic systems under arbitrary repulsive interactions.
Moreover, within this formalism, a common matrix is used for both the effective interactions projected onto Chern bands and the two-particle Hamiltonian, allowing the effective interactions to be characterized by the two-particle topology.
In this context, our hypothesis can be restated as follows: the Chern number defined for the effective interactions is related to the stabilization of FCIs.
To verify this hypothesis, we systematically investigate the relationship between two-particle Chern bands and the realization of the fermionic FCI at the filling factor $\nu = 1/3$ using exact diagonalization, and we find a certain degree of correlation.
From another perspective, our findings suggest that the topology inherent in the interaction itself can obstruct trivial ground states.
This indicates the possibility that considering various types of topological two-body interactions may lead to the realization of exotic states beyond FCIs.

This paper is organized as follows.
In Sec. \ref{conventionsection}, we introduce the notation used in this paper, as well as some subtle conventions that are often a source of confusion.
In particular, we introduce a matrix $\hat{S}_{\bm{q}}$ that characterizes both the effective interaction and the two-particle problem.
In Sec. \ref{sec:two-particle}, we formulate the two-particle problem using the matrix $\hat{S}_{\bm{q}}$.
In particular, we define the small Hamiltonian $\hat{h}_{\bm{q}}$, which possesses properties that are useful for discussing the topological characteristics of the system.
In Sec. \ref{sect:pseudo}, by comparing Haldane's pseudopotentials and two-particle Hamiltonians, we argue that there is a connection between the presence of dominant bound-state bands with a unit Chern number and the realization of an FCI.
In Sec. \ref{sec:numerical}, we systematically investigate the checkerboard lattice model using exact diagonalization to examine the correlation between the two-particle topology and the stability of the FCI at the filling factor $\nu = 1/3$.
In Sec. \ref{sec: obstruction}, we adopt a different perspective and explain how topological two-body interactions can obstruct the realization of trivial ground states.
Finally, in Sec. \ref{sec:conclusion}, we conclude the paper with a summary and discuss directions for future research.

\section{Convention and notation\label{conventionsection}}
In this section, we introduce the conventions and notations adopted in this paper.
In particular, we define a matrix $\hat{S}_{\bm{q}}$ that characterizes both the effective interaction and the two-particle problem.

\subsection{Non-interacting part}
As the free part of the system, we consider a quadratic fermionic lattice model given by
\begin{align}
    H_0=\sum_{\bm{R},\bm{R}'}\sum_{i,j}c^\dagger_{\bm{R},i}H_{(\bm{R},i),(\bm{R}',j)} c_{\bm{R}',j}, \label{tight}
\end{align}
where $(c,c^{\dagger})$ are annihilation and creation operators, $\bm{R} = (X, Y)$ labels the unit cell, and $i, j = 1, 2, \dots, n_{\mathrm{orb}}$ index the atomic orbitals within each unit cell.
Since our primary motivation lies in solid-state physics, we impose discrete translational symmetry on Eq.~(\ref{tight}).
As is well known, there are two common conventions for the Fourier transform in lattice systems \cite{vanderbilt2018berry}:
\begin{align}
    &c^{\dagger}_{\bm{k},i}=\frac{1}{\sqrt{N_{\rm unit}}}\sum_{\bm{R}}e^{i\bm{k}\cdot\bm{R}}c^{\dagger}_{\bm{R},i},\label{indep}\\
    &\tilde{c}^{\dagger}_{\bm{k},i}=\frac{1}{\sqrt{N_{\rm unit}}}\sum_{\bm{R}}e^{i\bm{k}\cdot \bm{R}+\bm{r}_i}c^{\dagger}_{\bm{R},i},\label{position-dependent}
\end{align}
where $\bm{k}$ denotes the crystal momentum, $N_{\rm unit}$ is the total number of unit cells, and $\bm{r}_i = (x_i, y_i)$ represents the intracell position of orbital $i$.
We adopt the convention (\ref{indep}). Under discrete translational symmetry, the Hamiltonian (\ref{tight}) can be rewritten as
\begin{align}
H_0=\sum_{\bm{k}}\sum_{i,j}c^\dagger_{\bm{k},i}[H_{\bm{k}}]_{i,j}
    c_{\bm{k},j}.
\end{align}
The one-particle dispersion is obtained by diagonalizing the Bloch Hamiltonian matrix $H_{\bm{k}}$:
\begin{align}
    H_{\bm{k}}=\sum_{\alpha}\epsilon_{\bm{k},\alpha}\ket{u_{\bm{k},\alpha}}\bra{u_{\bm{k},\alpha}},
\end{align}
where $\alpha$ is the band index, and $\ket{u_{\bm{k},\alpha}}$ denotes the periodic part of the Bloch eigenstate.
In the position basis $\{\ket{\bm{R},i}\}$, the Bloch eigenstates $\{\ket{\bm{k},\alpha}\}$ are represented as
\begin{align}
    \langle \bm{R},i\ket{\bm{k},\alpha}=u_{\bm{k},\alpha}(i)\frac{e^{i\bm{k}\cdot\bm{R}}}{\sqrt{N_{\rm unit}}},
\end{align}
where $u_{\bm{k},\alpha}(i):=\langle i\ket{u_{\bm{k},\alpha}}$. 
We assume that both $\ket{u_{\bm{k},\alpha}}$ and $\ket{\bm{k},\alpha}$ are normalized.

In addition to the Fourier transform, there also exists a convention ambiguity regarding the sign of the Berry connection in band theory.
We define the Berry connection, Berry curvature, and Chern number as
\begin{align}
    a_I(\bm{k},\alpha)&=-i\bra{u_{\bm{k}},\alpha}\partial_{k_{I}}\ket{u_{\bm{k}},\alpha},\\
    \omega(\bm{k},\alpha)&=\frac{\partial a_y(\bm{k},\alpha)}{\partial k_x}-\frac{\partial a_x(\bm{k},\alpha)}{\partial k_y},\\
    C_{\alpha}&=\int_{\rm BZ} \frac{d^2k}{2\pi}\omega(\bm{k},\alpha),
\end{align}
where $I = x, y$, and``BZ" denotes the Brillouin zone.
Note that the Berry connection and curvature can depend on the choice of Fourier transform convention \cite{simon2020contrasting}:
\begin{align}
    A_I(\bm{k},\alpha)&=-i\bra{U_{\bm{k}},\alpha}\partial_{k_{I}}\ket{U_{\bm{k}},\alpha}\notag\\
    &=a_I(\bm{k},\alpha)-\sum_{i}r^I_iu^*_{\bm{k}}(i)u_{\bm{k}}(i),\label{posi-berry}\\
    \Omega(\bm{k},\alpha)&=\frac{\partial A_y(\bm{k},\alpha)}{\partial k_x}-\frac{\partial A_x(\bm{k},\alpha)}{\partial k_y},
\end{align}
where $(r^{x}_i,r^{y}_i)=(x_i,y_i)$, and $\ket{U_{\bm{k},\alpha}}$ denotes the periodic part of a Bloch eigenstate in the convention (\ref{position-dependent}).
Nevertheless, topological numbers such as the Chern number, which are globally defined, are independent of the choice of convention \cite{simon2020contrasting}.

\subsection{Two-body interaction under projection}

As an interacting part, we consider the following repulsive Hamiltonian:
\begin{align}
    H_{\rm int}&=\frac{1}{2}\sum_{i,j,\bm{r},\bm{r}'}V^{ij}(\bm{r}-\bm{r}')c^{\dagger}_{\bm{r},i}c^{\dagger}_{\bm{r}',j}c_{\bm{r}',j}c_{\bm{r},i}\notag\\
    &=\frac{1}{2N_{\rm unit}}\sum_{i,j}\sum_{\bm{q},\bm{k},\bm{k}',\Delta\bm{r}}e^{i\bm{q}\cdot(\Delta\bm{r})}V^{ij}(\Delta\bm{r})\notag\\
    &~~~~c^{\dagger}_{\bm{k},i}c^{\dagger}_{\bm{k}',j}c_{\bm{k}'-\bm{q},j}c_{\bm{k}+\bm{q},i}\notag\\
    &=\frac{1}{2N_{\rm unit}}\sum_{i,j}\sum_{\bm{q},\bm{k}_1,\bm{k}_2,\Delta\bm{r}}e^{-i(\bm{k}_1-
    \bm{k}_2)\cdot\Delta\bm{r}}V^{ij}(\Delta\bm{r})\notag\\
    &~~~~c^{\dagger}_{\bm{k}_1,i}c^{\dagger}_{\bm{q}-\bm{k}_1,j}c_{\bm{q}-\bm{k}_2,j}c_{\bm{k}_2,i},
\end{align}
where $\Delta\bm{r}:=\bm{r}-\bm{r}'$.
We consider the case where there is a single occupied band of interest, and the gap to the other bands is much larger than the interaction strength. 
Furthermore, we assume that the interaction strength is much larger than the bandwidth of the lowest band $\alpha$.
In such a situation, the low-energy physics of the Hamiltonian $H_0+H_{\rm int}$ is well described by the following projected Hamiltonian:
\begin{align}
    \mathcal{P}_{\alpha}H_{\rm int}\mathcal{P}_{\alpha}
    =&\frac{1}{2N_{\rm unit}}\sum_{i,j}\sum_{\bm{q},\bm{k}_1,\bm{k}_2,\Delta\bm{r}}e^{-i(\bm{k}_1-\bm{k}_2)\cdot\Delta\bm{r}}V^{ij}(\Delta\bm{r})\notag\\
    &u^{*}_{\bm{k}_1,\alpha}(i)u^{*}_{\bm{q}-\bm{k}_1,\alpha}(j)u_{\bm{q}-\bm{k_2},\alpha}(j)u_{\bm{k}_2,\alpha}(i)\notag\\
    &c^{\dagger}_{\bm{k}_1,\alpha}c^{\dagger}_{\bm{q}-\bm{k}_1,\alpha}c_{\bm{q}-\bm{k}_2,\alpha}c_{\bm{k}_2,\alpha}.
\end{align}
\subsection{Two-particle orbitals and symmetrized Hamiltonian}
Up to lattice translations, the two-body interaction is determined by the relative positions of the two particles specified by the following set:
\begin{align}
    a:=\{i,j,\Delta\bm{r}\}.
\end{align}
Physically, this information describes the degrees of freedom for the relative motion relevant to the interaction Hamiltonian.
We refer to this set as the two-particle orbital.

To avoid redundancy, we identify the pairs $\{i,j,\Delta\bm{r}\}$ and $\{j,i,-\Delta\bm{r}\}$, which contribute equally to $H_{\rm int}$. Consequently, we omit the factor of $1/2$ in the interacting Hamiltonian.
For convenience, we define a matrix $\hat{A}_{\bm{q}}$ whose elements are given by
\begin{align}
    [\hat{A}_{\bm{q}}]_{a,\bm{k}}=\sqrt{\frac{V^{ij}(\Delta\bm{r})}{N_{\rm unit}}}e^{i\bm{k}\cdot\Delta\bm{r}}u_{\bm{k},\alpha}(i)u_{\bm{q}-\bm{k},\alpha}(j).\label{amatrix}
\end{align}
Using this matrix, the projected Hamiltonian can be rewritten as
\begin{align}
    \mathcal{P}_{\alpha}H_{\rm int}\mathcal{P}_{\alpha}=\sum_{\bm{k}_1,\bm{k}_2,\bm{q}}[\hat{A}^{\dagger}_{\bm{q}}\hat{A}_{\bm{q}}]_{\bm{k}_1,\bm{k}_2}c^{\dagger}_{\bm{k}_1}c^{\dagger}_{\bm{q}-\bm{k}_1}c_{\bm{q}-\bm{k}_2}c_{\bm{k}_2}.
\end{align}
In the summation over the matrix index, we identify the pairs $\{i,j,\Delta\bm{r}\}$ and $\{j,i,-\Delta\bm{r}\}$ and omit the factor $1/2$ in the interacting Hamiltonian.
We omit the index $\alpha$ henceforth.
For convenience, we antisymmetrize the Hamiltonian matrix elements:
\begin{align}
    \mathcal{P}_{\alpha}H_{\rm int}\mathcal{P}_{\alpha}&=\frac{1}{2}\sum_{\bm{k}_1,\bm{k}_2,\bm{q}}[\hat{S}^{\dagger}_{\bm{q}}\hat{S}_{\bm{q}}]_{\bm{k}_1,\bm{k}_2}c^{\dagger}_{\bm{k}_1}c^{\dagger}_{\bm{q}-\bm{k}_1}c_{\bm{q}-\bm{k}_2}c_{\bm{k}_2},\\
    [\hat{S}_{\bm{q}}]_{a,\bm{k}}&=\frac{[\hat{A}_{\bm{q}}]_{a,\bm{k}}-[\hat{A}_{\bm{q}}]_{a,\bm{q}-\bm{k}}}{\sqrt{2}}.\label{scatteringelement}
\end{align}
The fermionic antisymmetry is encoded into the following relation:
\begin{align}
    [\hat{S}_{\bm{q}}]_{a,\bm{k}}=-[\hat{S}_{\bm{q}}]_{a,\bm{q}-\bm{k}}~.\label{anti-symmetry}
\end{align}

\section{Two-particle problem\label{sec:two-particle}}
In this section, we develop a theoretical framework for the two-particle problem using the matrix $\hat{S}_{\bm{q}}$, which appears in the definition of the effective interaction (\ref{scatteringelement}).
We introduce the large and small Hamiltonians, which share the same nonzero eigenenergies.
These two Hamiltonians are related through the singular value decomposition of $\hat{S}_{\bm{q}}$.
To provide an intuitive understanding, we also introduce the real-space hopping Hamiltonian for the bound states, which is given by the Fourier transform of the small Hamiltonian.
All of these results can be straightforwardly extended to bosons, and some subtle points are discussed in Appendix.
The two-particle problem in bosonic and fermionic systems is widely formulated in various contexts. For example, see Refs. \cite{Valiente-Petrosyan-08, Salerno-18,Torma2018,Iskin2021,Iskin2023,Ling-20}.

\subsection{Large and small Hamiltonian}

The following basis spans the two-particle Hilbert space:
\begin{align}
    \ket{\bm{q};\bm{k}}:=c^{\dagger}_{\bm{q}-\bm{k}}c^{\dagger}_{\bm{k}}\ket{0}~~\mathrm{for}~n(\bm{k})< n(\bm{q}-\bm{k}),
\end{align}
where $n(\bm{k})$ denotes the index when the momenta are arranged in a suitable order.
Using the Bloch-De Dominicis theorem, we obtain
\begin{align}
    &\bra{\bm{q};\bm{k}}c^{\dagger}_{\bm{k}_1}c^{\dagger}_{\bm{q}-\bm{k}_1}c_{\bm{q}-\bm{k}_2}c_{\bm{k}_2}
    \ket{\bm{q};\bm{k}'}\notag\\
    &=\bra{0}c_{\bm{k}}c_{\bm{q}-\bm{k}}c^{\dagger}_{\bm{k}_1}c^{\dagger}_{\bm{q}-\bm{k}_1}c_{\bm{q}-\bm{k}_2}c_{\bm{k}_2}c^{\dagger}_{\bm{q}-\bm{k}'}c^{\dagger}_{\bm{k}'}\ket{0}\notag\\
    &=[\delta_{\bm{k},\bm{k}_1}-\delta_{\bm{k},\bm{q}-\bm{k}_1}][\delta_{\bm{k}',\bm{k}_2}-\delta_{\bm{k}',\bm{q}-\bm{k}_2}].
\end{align}
Then, the matrix representation of the two-particle Hamiltonian is given by
\begin{align}
&[\hat{\mathcal{H}}_{\bm{q}}]_{\bm{k},{\bm{k}'}}:=\bra{\bm{q};\bm{k}}\mathcal{P}_{\alpha}H_{\rm int}\mathcal{P}_{\alpha}\ket{\bm{q};\bm{k}'}=2~[\hat{S}^{\dagger}_{\bm{q}}\hat{S}_{\bm{q}}]_{\bm{k},\bm{k}'}.\label{largeham}
\end{align}
We refer to this Hamiltonian as the ``large Hamiltonian".
Instead of this Hamiltonian, we consider the ``small" Hamiltonian, which shares the same nonzero eigenvalues:
\begin{align}
    &[\hat{h}_{\bm{q}}]_{a,a'}:=\sum_{\{\bm{k}|n(\bm{k})<n(\bm{q}-\bm{k})\}}2[\hat{S}_{\bm{q}}]_{a,\bm{k}}[\hat{S}^{\dagger}_{\bm{q}}]_{\bm{k},a'}\notag\\
    =&\sum_{\{\bm{k}|n(\bm{k})<n(\bm{q}-\bm{k})\}}[\hat{S}_{\bm{q}}]_{a,\bm{k}}[\hat{S}^{\dagger}_{\bm{q}}]_{\bm{k},a'}
    +[\hat{S}_{\bm{q}}]_{a,\bm{q}-\bm{k}}[\hat{S}^{\dagger}_{\bm{q}}]_{\bm{q}-\bm{k},a'}\notag\\
    =&\sum_{\{\bm{k}|n(\bm{k})<n(\bm{q}-\bm{k})\}}[\hat{S}_{\bm{q}}]_{a,\bm{k}}[\hat{S}^{\dagger}_{\bm{q}}]_{\bm{k},a'}
    \notag\\
    +&\sum_{\{\bm{k}|n(\bm{k})>n(\bm{q}-\bm{k})\}}[\hat{S}_{\bm{q}}]_{a,\bm{k}}[\hat{S}^{\dagger}_{\bm{q}}]_{\bm{k},a'}\notag\\
    =&\sum_{\bm{k}}[\hat{S}_{\bm{q}}]_{a,\bm{k}}[\hat{S}^{\dagger}_{\bm{q}}]_{\bm{k},a'},\label{smallhamexpression}
\end{align}
where the final summation is taken over the entire Brillouin zone.
We have used the property (\ref{anti-symmetry}). Note that $[\hat{S}_{\bm{q}}]_{a,\bm{k}}=0$ for $n(\bm{k})=n(\bm{q}-\bm{k})$.
Although there are some differences, a similar formulation can be made in the case of bosons as well.
This formulation can also be extended to the cases where there are multiple target bands for the projection.
See Appendix for details.

The factor of 2 in Eq. (\ref{largeham}) arises from the limitation on $\bm{k}$, $n(\bm{k})<n(\bm{q}-\bm{k})$.
The complete form of the matrix $\hat{S}^{\dagger}_{\bm{q}}\hat{S}_{\bm{q}}$ has the following relationship with the matrix $\hat{\mathcal{H}}_{\bm{q}}/2$ if there is no momentum $\bm{k}$ satisfying $n(\bm{k})=n(\bm{q}-\bm{k})$:
\begin{align}
    \hat{S}^{\dagger}_{\bm{q}}\hat{S}_{\bm{q}}=
    \begin{pmatrix}
        \hat{\mathcal{H}}_{\bm{q}}/2&-\hat{\mathcal{H}}_{\bm{q}}/2\\
        -\hat{\mathcal{H}}_{\bm{q}}/2&\hat{\mathcal{H}}_{\bm{q}}/2
    \end{pmatrix}
    =
    \begin{pmatrix}
        1/2&-1/2\\
        -1/2&1/2
    \end{pmatrix}
    \otimes \hat{\mathcal{H}}_{\bm{q}}.
\end{align}
Thus, all the nonzero eigenvalues of $\hat{S}^{\dagger}_{\bm{q}}\hat{S}_{\bm{q}}$ are equal to the corresponding eigenvalues of $\hat{\mathcal{H}}_{\bm{q}}$.
The presence of a vector $\bm{k}$ satisfying $n(\bm{k})=n(\bm{q}-\bm{k})$ simply adds one column vector and one row vector of zeros and does not affect the discussion of the nonzero eigenvalues.

The number of the nonzero eigenvalues of the two-particle large and small Hamiltonian is at most the size of the matrix $\hat{h}_{\bm{q}}=\hat{S}_{\bm{q}}\hat{S}^{\dagger}_{\bm{q}}$, or the number of the two-particle orbitals $\{a=\{i,j,\Delta\bm{r}\}\}$.
Physically, these eigenstates correspond to bound states of two particles.
If $V^{ij}(\Delta\bm{r})$ is a finite-range interaction, the number of nonzero eigenvalues is finite, and the majority of the eigenvalues of the large Hamiltonian are zero.
Such zero modes correspond to scattering states, each composed of two free particles.

The small Hamiltonian $\hat{h}_{\bm{q}}=\hat{S}_{\bm{q}}\hat{S}^{\dagger}_{\bm{q}}$ has the advantage of being invariant under the following gauge transformation:
\begin{align}
    \bm{u}_{\bm{k},\alpha}\rightarrow e^{i\chi(\bm{k})}\bm{u}_{\bm{k},\alpha}.
\end{align}
Under this transformation, $ [\hat{S}_{\bm{q}}]_{a,\bm{k}}$ transforms as $ e^{i[\chi(\bm{k})+\chi(\bm{q}-\bm{k})]}[\hat{S}_{\bm{q}}]_{a,\bm{k}}$, but this change does not affect the summation over momentum in the calculation of the matrix product of $\hat{S}_{\bm{q}}\hat{S}^{\dagger}_{\bm{q}}$.
Moreover, under the convention (\ref{indep}), $\bm{u}_{\bm{k},\alpha}$ is continuous and periodic in the Brillouin zone, except for the gauge degrees of freedom, which ensures that $\hat{h}_{\bm{q}}$ is also continuous and periodic in $\bm{q}$.
These properties are essential for considering the topology of the two-particle band structure.
On the other hand, the above gauge transformation changes the matrix elements of the large Hamiltonian, implying that it is not gauge invariant and, in general, not continuous.

\subsection{Spectral and singular value decompositions}
The bound-state energies are given by the nonzero eigenvalues of $\hat{S}^{\dagger}_{\bm{q}}\hat{S}_{\bm{q}}$ or $\hat{S}_{\bm{q}}\hat{S}^{\dagger}_{\bm{q}}$.
To gain further insight, it is instructive to introduce the singular value decomposition.
We begin with the spectral decomposition of $\hat{S}^{\dagger}_{\bm{q}}\hat{S}_{\bm{q}}$:
\begin{align}
&\hat{S}^{\dagger}_{\bm{q}}\hat{S}_{\bm{q}}=\sum_{\beta}\ket{\Psi_{\bm{q},\beta}}\epsilon_{\bm{q},\beta}\bra{\Psi_{\bm{q},\beta}}\notag\\
&=\sum_{\beta,\bm{r},a}\ket{\Psi_{\bm{q},\beta}}\sqrt{\epsilon_{\bm{q},\beta}}\bra{\Psi_{\bm{q},\beta}}\bm{r},a\rangle
    \bra{\bm{r},a} \Psi_{\bm{q},\beta}\rangle \sqrt{\epsilon_{\bm{q},\beta}}\langle \Psi_{\bm{q},\beta}|\notag\\
&=\sum_{\beta,a}\ket{\Psi_{\bm{q},\beta}}\sqrt{\epsilon_{\bm{q},\beta}}\bra{\psi_{\bm{q},\beta}}a\rangle
    \bra{a} \psi_{\bm{q},\beta}\rangle \sqrt{\epsilon_{\bm{q},\beta}}\langle \Psi_{\bm{q},\beta}|.\label{spectraldecomposition}
\end{align}
Here, we have defined the two-particle Bloch wave function $\bra{\bm{r},a} \Psi_{\bm{q},\beta}\rangle$ and its periodic part $\bra{a} \psi_{\bm{q},\beta}\rangle$, which satisfies the following Bloch theorem:
\begin{align}
    \bra{\bm{r},a} \Psi_{\bm{q},\beta}\rangle =\frac{e^{i\bm{q}\cdot\bm{r}}}{\sqrt{N_{\rm unit}}}\bra{a} \psi_{\bm{q},\beta}\rangle.
\end{align}
In this notation, the positions of two particles are given by $\{\bm{r},i\}$ and $\{\bm{r}-\Delta\bm{r},j\}$. 
The states $|\psi_{\bm{q},\beta}\rangle$ are nothing but the eigenstates of the small Hamiltonian.
This identification becomes evident by introducing the singular value decomposition of $\hat{S}_{\bm{q}}$:
\begin{align}
    &\hat{S}_{\bm{q}}=
    \hat{U}_{\bm{q}}\hat{\Sigma}_{\bm{q}}\hat{V}^{\dagger}_{\bm{q}},\label{svd}
\end{align}
where $\hat{U}_{\bm{q}}$ and $\hat{V}_{\bm{q}}$ are unitary matrices, and $\hat{\Sigma}_{\bm{q}}$ is a positive semidifinite diagonal matrix.
By comparing Eqs.(\ref{spectraldecomposition}) and (\ref{svd}), we obtain the following representations:
\begin{align}
    &[\hat{U}_{\bm{q}}]_{a,\beta}=\bra{a} \psi_{\bm{q},\beta}\rangle,\\
    &[\hat{\Sigma}_{\bm{q}}]_{\beta,\beta'}=\delta_{\beta,\beta'}\sqrt{\epsilon_{\bm{q},\beta}},\\
    &[\hat{V}_{\bm{q}}^{\dagger}]_{\beta,\bm{k}}=\langle\Psi_{\bm{q},\beta}\ket{\bm{q};\bm{k}},
\end{align}
where the definition of $\{\ket{\bm{q};\bm{k}}\}$ is extended to $\bm{k}$ in the whole Brillouin zone.
Using these relations, the spectral decomposition of the small Hamiltonian becomes
\begin{align}
    \hat{h}_{\bm{q}}&=\hat{S}_{\bm{q}}\hat{S}^{\dagger}_{\bm{q}}=\hat{U}_{\bm{q}}\hat{\Sigma}_{\bm{q}}\hat{V}^{\dagger}_{\bm{q}}\hat{V}_{\bm{q}}\hat{\Sigma}_{\bm{q}}\hat{U}^{\dagger}_{\bm{q}}=\hat{U}_{\bm{q}}\hat{\Sigma}^2_{\bm{q}}\hat{U}^{\dagger}_{\bm{q}}\notag\\
    &=\sum_{\beta}\ket{\psi_{\bm{q},\beta}}\epsilon_{\bm{q},\beta}\bra{\psi_{\bm{q},\beta}}.
\end{align}
This equation shows that the states $\ket{\psi_{\bm{q},\beta}}$ are the eigenstates of the small Hamiltonian.
Correspondingly, the total bound-state Hamiltonian is written in terms of the two-particle Bloch wave functions:
\begin{align}
    \bra{\bm{r},a}\hat{h}^{(2p)}\ket{\bm{r}',a'}&=
    \sum_{\bm{q},\beta}\bra{\bm{r},a} \Psi_{\bm{q},\beta}\rangle \epsilon_{\bm{q},\beta} \bra{\Psi_{\bm{q},\beta}}\bm{r}',a'\rangle\notag\\
    &=\sum_{\bm{q},\beta}\Psi_{\bm{q},\beta}(\bm{r},a)\epsilon_{\bm{q},\beta}\Psi^*_{\bm{q},\beta}(\bm{r}',a').\label{two-particle-elemetns}
\end{align}
Note that $\bm{r}$ and $a$ describe the position of the two-particle orbital and the relative position of the two particles, respectively.
To avoid redundancy, we have identified $\{j,i,-\Delta\bm{r}\}$ with $\{i,j,\Delta\bm{r}\}$. 
If  we were to formulate the same formalism for $\{j,i,-\Delta\bm{r}\}$, the following relation would hold:
\begin{align}
    [\hat{S}_{\bm{q}}]_{\{ j,i,-\Delta\bm{r}\},\bm{k}}=-e^{-i\bm{q}\cdot\Delta\bm{r}} [\hat{S}_{\bm{q}}]_{\{ i,j,\Delta\bm{r}\},\bm{k}}.
\end{align}
With an appropriate phase factor, the following antisymmetry with respect to the relative position holds:
\begin{align}
    \Psi_{\bm{q},\beta}(\bm{r},\{ j,i,-\Delta\bm{r}\})&=-e^{-i\bm{q}\cdot\Delta\bm{r}} \Psi_{\bm{q},\beta}(\bm{r},\{ i,j,\Delta\bm{r}\})
    \notag\\
    &=-\Psi_{\bm{q},\beta}(\bm{r}-\Delta\bm{r},\{ i,j,\Delta\bm{r}\}).
\end{align}

\subsection{Real-space picture}
The small Hamiltonian describes the bound-state Hamiltonian whose subscript is a two-particle orbital.
In this paper, we primarily work in momentum space. However, in this subsection, we derive the real-space expression of the small Hamiltonian to gain deeper physical intuition.
In particular, we decompose the two-particle hopping
into products of one-particle correlation functions in real space.
Instead of $\hat{S}_{\bm{q}}$, we use $\hat{A}_{\bm{q}}$:
\begin{align}
    [\hat{h}_{\bm{q}}]_{a,a'}=\sum_{\bm{k}}[\hat{A}_{\bm{q}}]_{a,\bm{k}}[\hat{A}^{\dagger}_{\bm{q}}]_{\bm{k},a'}-[\hat{A}_{\bm{q}}]_{a,\bm{k}}[\hat{A}^{\dagger}_{\bm{q}}]_{\bm{q-}\bm{k},a'}.
\end{align}
We set $a,a'=\{i,j,\Delta\bm{r}\},\{k,l,\Delta\bm{r}'\}$.
Then, the first term becomes
\begin{align}
    &\sum_{\bm{k}}[\hat{A}_{\bm{q}}]_{\{i,j,\Delta\bm{r}\},\bm{k}}[\hat{A}^{\dagger}_{\bm{q}}]_{\bm{k},\{k,l,\Delta\bm{r}'\}}=\sqrt{V^{ij}(\Delta\bm{r})V^{kl}(\Delta\bm{r}')}\notag\\
    &\frac{1}{N_{\rm unit}}\sum_{\bm{k}}e^{i\bm{k}\cdot(\Delta\bm{r}-\Delta\bm{r}')}[\hat{P}_{\bm{k}}]_{ik}[\hat{P}_{\bm{q}-\bm{k}}]_{jl},
\end{align}
where 
\begin{align}
    \hat{P}_{\bm{k}}=\ket{u_{\bm{k}}}\bra{u_{\bm{k}}}
\end{align}
is the projection operator onto the target band.
Similarly, the second term becomes
\begin{align}
    &\sum_{\bm{k}}[\hat{A}_{\bm{q}}]_{\{i,j,\Delta\bm{r}\},\bm{k}}[\hat{A}^{\dagger}_{\bm{q}}]_{\bm{q}-\bm{k},\{k,l,\Delta\bm{r}'\}}=\sqrt{V^{ij}(\Delta\bm{r})V^{kl}(\Delta\bm{r}')}\notag\\
    &\frac{1}{N_{\rm unit}}\sum_{\bm{k}}e^{i\bm{k}\cdot\Delta\bm{r}}[\hat{P}_{\bm{k}}]_{il}e^{-i(\bm{q}-\bm{k})\cdot\Delta\bm{r}'}[\hat{P}_{\bm{q}-\bm{k}}]_{jk}.
\end{align}
\begin{figure}[]
\begin{center}
 \includegraphics[width=8cm,angle=0,clip]{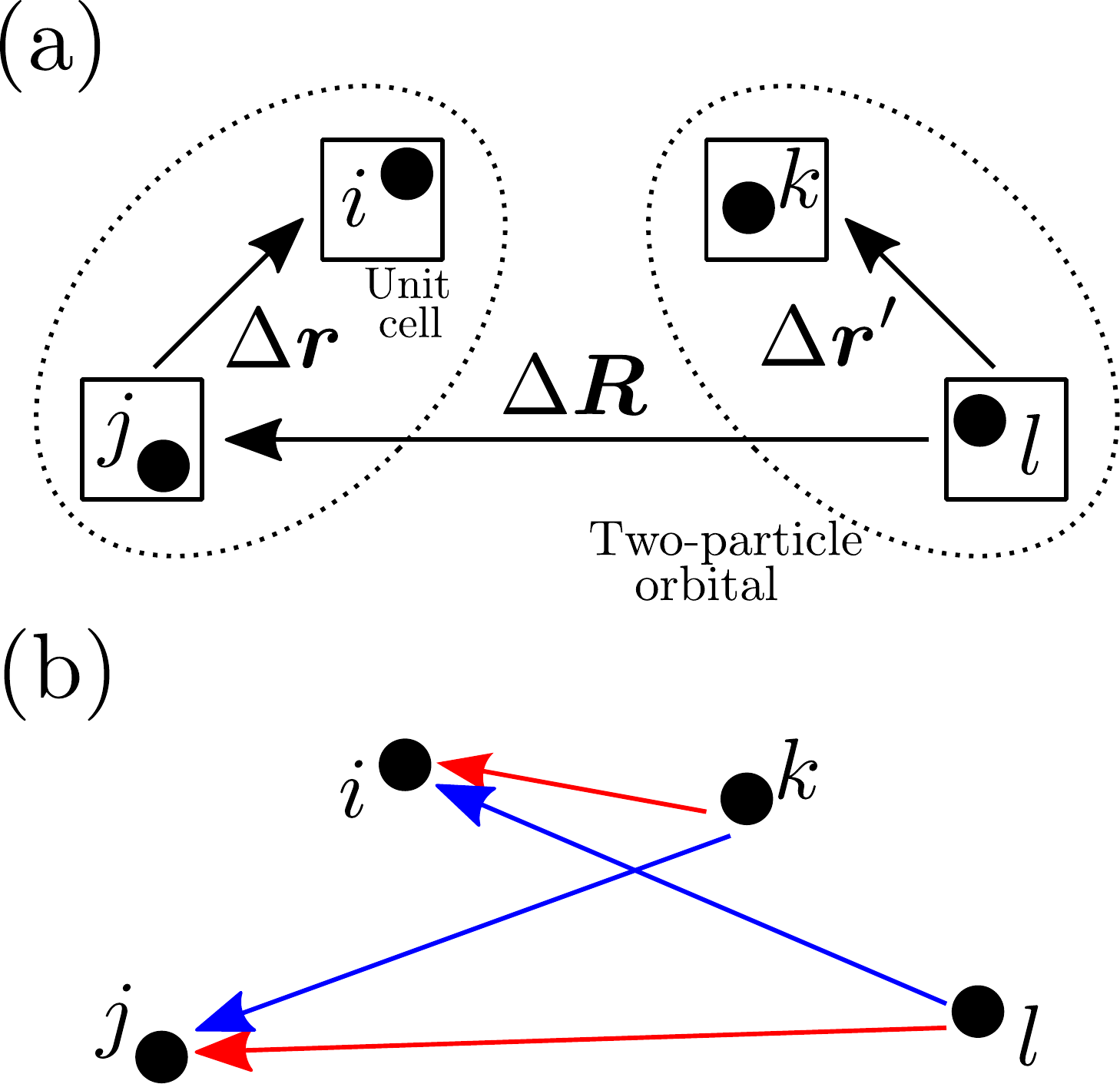}
 \caption{Real-space pictures of (a) two-particle hopping and (b) fermionic exchange process.} 
 \label{fig1}
\end{center}
\end{figure}
The two-particle hopping Hamiltonian is given by the Fourier transform:
\begin{align}
    [\hat{h}({\Delta\bm{R}})]_{a,a'}
    :=&\frac{1}{N_{\rm unit}}\sum_{\bm{q}}e^{i\bm{q}\cdot\Delta\bm{R}}[\hat{h}_{\bm{q}}]_{a,a'}\notag\\
    =&\sqrt{V^{ij}(\Delta\bm{r})}\sqrt{V^{kl}(\Delta\bm{r}')}\notag\\
    &\left[[\hat{P}(\Delta\bm{R}+\Delta\bm{r}-\Delta\bm{r}')]_{ik}[\hat{P}(\Delta\bm{R})]_{jl}\right.\notag\\
    &\left.-[\hat{P}(\Delta\bm{R}+\Delta\bm{r})]_{il}[\hat{P}(\Delta\bm{R}-\Delta\bm{r}')]_{jk}\right].
\end{align}
Here, we define the Fourier transform of $\hat{P}_{\bm{k}}$:
\begin{align}
    [\hat{P}(\Delta\bm{R})]_{ij}:=\frac{1}{N_{\rm unit}}\sum_{\bm{k}}e^{i\bm{k}\cdot\Delta\bm{R}}[\hat{P}_{\bm{k}}]_{ij},
\end{align}
which is identical to the matrix representation of the projection operator onto the Chern band (i.e., the one-particle correlation function):
\begin{align}
    &\bra{\bm{R}+\Delta\bm{R},i}\hat{P}_{\alpha}\ket{\bm{R},j}\notag\\
    =&\sum_{\bm{k}}\frac{1}{\sqrt{N_{\rm unit}}}u_{\bm{k},\alpha}(i)e^{i\bm{k}\cdot(\bm{R}+\Delta\bm{R})}
    \frac{1}{\sqrt{N_{\rm unit}}}u^*_{\bm{k},\alpha}(j)e^{-i\bm{k}\cdot\bm{R}}\notag\\
    =&\frac{1}{N_{\rm unit}}\sum_{\bm{k}}e^{i\bm{k}\cdot\Delta\bm{R}}[\hat{P}_{\bm{k}}]_{ij}=[\hat{P}(\Delta\bm{R})]_{ij}.
\end{align}
In summary, the two-particle hopping is decomposed into one-particle propagation within the band $\alpha$, taking into account the fermionic antisymmetry (Fig. \ref{fig1}).

\section{Pseudopotential versus
Two-particle topological bands\label{sect:pseudo}}
In this section, we argue that the topology of the two-particle bands — or equivalently, the topology of the effective interaction — plays a crucial role in mimicking the pseudopotentials known from FQHE physics.
We begin by considering the two-particle problem on the lowest Landau level.
Let us consider an interaction that depends only on the relative distance between two particles.
The two-particle Hamiltonian in the lowest Landau level is expressed in terms of the relative/center-of-mass angular momenta $m$ and $M$:
\begin{align}
    h^{(2p)}_{\rm LLL}=\sum_{m,M}\ket{m,M}V_m\bra{m,M}.
\end{align}
Due to rotational symmetry, the eigenvalues depend only on $m$.
These eigenvalues, denoted by $\{V_m\}$, are known as Haldane's pseudopotentials \cite{Haldane-Hierarchy-83}. 
In the fermionic (bosonic) case, $m$ is restricted to odd (even) integers due to the required antisymmetry (symmetry) of the wavefunction.
The corresponding two-body interaction leads to the exact FQHE ground states at fractional fillings \cite{yoshioka-textbook-02}.
In general, the fermionic (bosonic) Laughlin state at filling fraction $\nu=1/(2n+1)$ ($\nu=1/2n$) can be realized by including the pseudopotentials $\{V_m\}$ up to $m=2n-1$ $(m=2n-2)$.

For simplicity, we fix $m$ to a specific value.
The matrix elements in the real-space basis are
\begin{align}
    \bra{Z,z}h^{(2p)}_{\rm LLL}\ket{Z',z'}&=\sum_{M}\langle Z,z\ket{m,M}V_m\bra{m,M} Z',z'\rangle\notag\\
    &=\sum_{M}\Psi_{M,m}(Z,z)V_m\Psi^*_{M,m}(Z',z'),\label{two-particle-landau-problem}
\end{align}
where $(Z,z)=\left(\frac{z_1+z_2}{2},z_1-z_2\right)$ are the center-of-mass and relative complex coordinates, respectively.
The explicit form of the two-particle wave function $\Psi_{M,m}(Z,z)$ is given in Ref. \cite{yoshioka-textbook-02} as 
\begin{align}
    \Psi_{M,m}(Z,z)\propto Z^Mz^m \exp\left[-\frac{|z|^2}{8} -\frac{|Z|^2}{2}\right].\label{two-particle-landau}
\end{align}

Clearly, $M,m$ correspond to $\bm{q},\beta$ in the lattice two-particle problem discussed above.
For a fixed $m$, the center-of-mass part of the wavefunctions forms the two-particle Landau level.
In our previous work \cite{okuma2023relationship}, we predicted that the two-particle topology of the bosonic bound states is an important factor for the realization of the bosonic FCI in models with the onsite repulsive interaction.
In our present formalism, this point becomes explicitly evident by comparing Eqs.(\ref{two-particle-elemetns}) and (\ref{two-particle-landau-problem}), as will be discussed below.
Furthermore, it is worth emphasizing that by switching the symmetry of $\hat{S}_{\bm{q}}$, one can treat both bosonic and fermionic models with arbitrary interactions within a unified framework. 

The two-particle orbital $a$, or $\{i,j,\Delta\bm{r}\}$, describes the relative coordinate relevant to the interaction. The unit-cell vector $\bm{r}$ describes the position of the two-particle orbital, which corresponds to the center-of-mass coordinate in the Landau level problem. 
Owing to this correspondence, the set of two-particle Bloch eigenstates $\{\Psi_{\bm{q},\beta}(\bm{r},a)\}$ with the nonzero eigenenergies forms the two-particle Chern bands with the unit Chern number. Reference \cite{Lauchli-Liu-Bergholtz-Moessner-13} pointed out that a single pseudopotential $V_m$ is expected to correspond to the two two-particle bands. This is because the sum of two crystal momenta spans a region with twice the area of the Brillouin zone.
In our topological interpretation, this implies that the sum of the Chern numbers of the two highest-energy two-particle bands equals unity.
Since the total Chern number of all bands in the small Hamiltonian is zero, this correspondence requires at least three bound-state bands.
In other words, the number of two-particle orbitals $a=\{i,j,\Delta\bm{r}\}$ must be at least three, for the correspondence with pseudopotentials.
Since the definition of the effective interaction (\ref{scatteringelement}) and that of the small Hamiltonian use the same matrix $\hat{S}_{\bm{q}}$, the above hypothesis can be restated as follows: the Chern number of the effective interaction is related to the stabilization of the FCI.

Note, however, that the nontrivial two-particle topology is not sufficient to guarantee full correspondence with the FCI stabilization.
Under the singular value decomposition (\ref{svd}),
the small Hamiltonian $\hat{h}_{\bm{q}}$ and the scattering matrix $\hat{S}^{\dagger}_{\bm{q}}\hat{S}_{\bm{q}}$ in the interaction Hamiltonian (\ref{scatteringelement}) are expressed as
\begin{align}
    \hat{h}_{\bm{q}}&=\hat{S}_{\bm{q}}\hat{S}^{\dagger}_{\bm{q}}=\hat{U}_{\bm{q}}\hat{\Sigma}^2_{\bm{q}}\hat{U}^{\dagger}_{\bm{q}},\\
    \hat{S}^{\dagger}_{\bm{q}}\hat{S}_{\bm{q}}&=\hat{V}_{\bm{q}}\hat{\Sigma}^2_{\bm{q}}\hat{V}^{\dagger}_{\bm{q}}.
\end{align}
While $\hat{U}_{\bm{q}}$ contains the topological information, the many-body scattering is determined by $\hat{V}_{\bm{q}}$.
Since $\hat{S}_{\bm{q}}$ and $\hat{\Sigma}_{\bm{q}}$ are not arbitrary matrices, $\hat{V}_{\bm{q}}$ is not independent of $\hat{U}_{\bm{q}}$.
However, the correspondence between $\hat{V}_{\bm{q}}$ and the pseudopotentials is indirect.
In this sense, the topology of $\hat{h}_{\bm{q}}$ reflects only one aspect of the effective interaction. Therefore, one cannot conclude the realization of FCI ground states based solely on the two-particle topology.
Nevertheless, the two-particle topology remains a useful indicator in the search for FCIs.

\section{Numerical calculations\label{sec:numerical}}
\begin{figure}[]
\begin{center}
 \includegraphics[width=7.5cm,angle=0,clip]{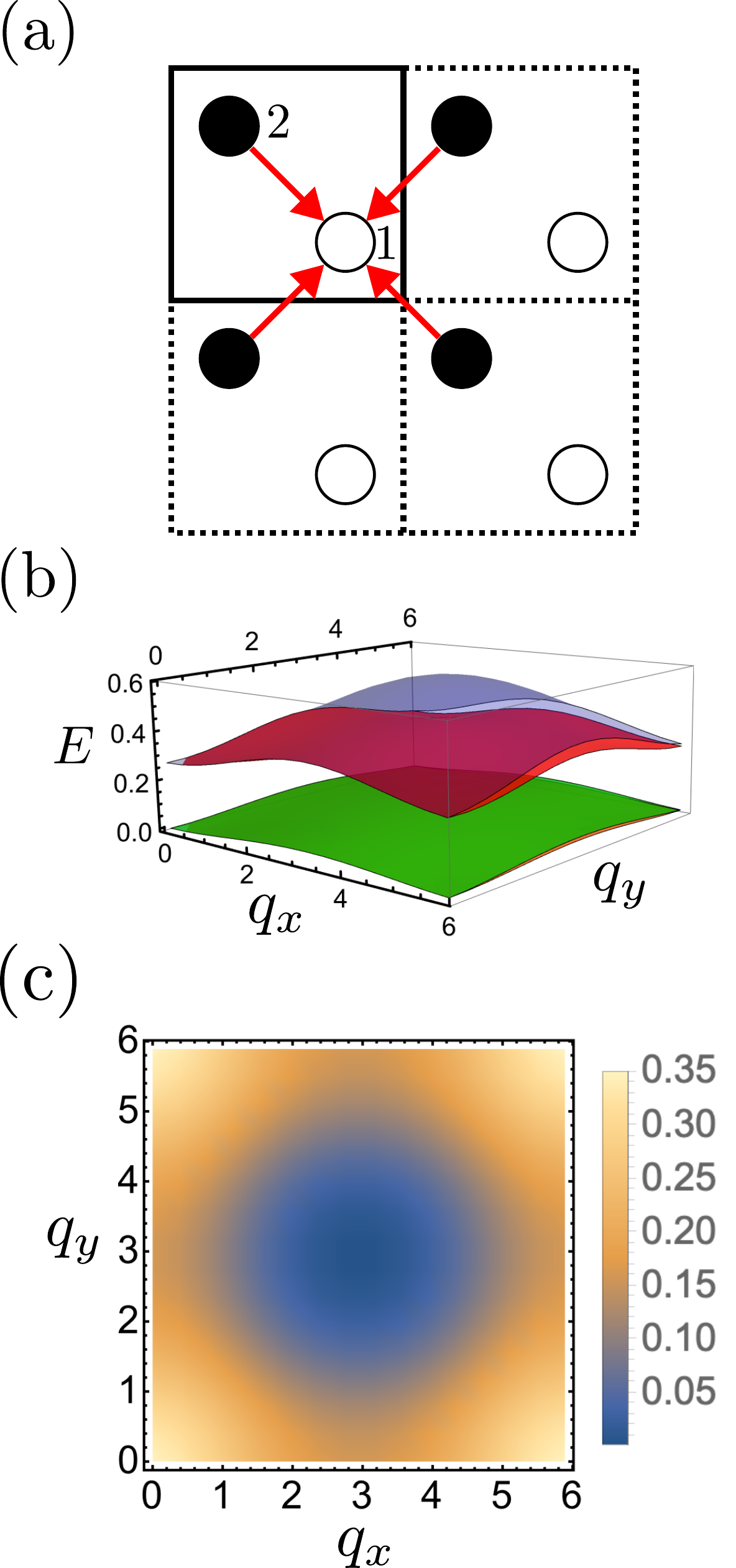}
 \caption{Two-particle problem in the checkerboard lattice model. (a) Unit-cell structure. Red arrows represent two-particle orbitals. (b) Band structure of the bound states. (c) Berry curvature of the top two bands. } 
 \label{fig2}
\end{center}
\end{figure}

\begin{figure*}[]
\begin{center}
 \includegraphics[width=16cm,angle=0,clip]{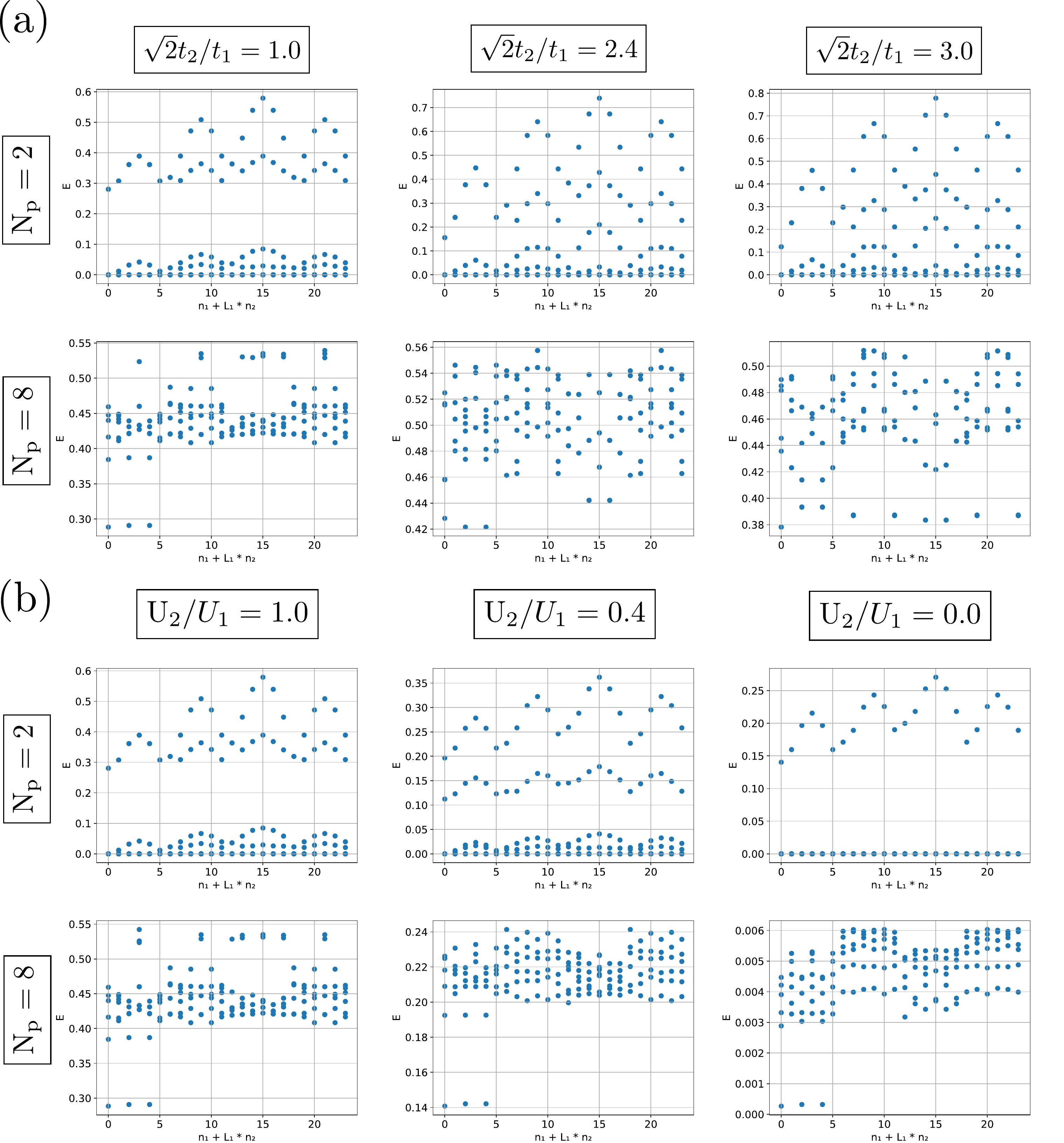}
 \caption{Exact Diagonalizations of the checkerboard lattice model under parameter variation. The system size is $L_1\times L_2=6\times4$. Total momentum sectors are labeled by $(n_1,n_2)$. (a) Hopping dependence of two- and eight-particle spectra. (b) Interaction dependence of two- and eight-particle spectra.} 
 \label{fig3}
\end{center}
\end{figure*}

In this section, we perform numerical calculations for the checkerboard lattice model to investigate the relationship between two-particle topology and the FCI stabilization.
This model was studied in the context of the fermionic FCI \cite{Neupert-Santos-Chamon-Mudry-11,Sun-Gu-Katsura-DasSarma-11,sheng2011fractional}. 
The two-particle problem for this model was studied in Ref. \cite{Lauchli-Liu-Bergholtz-Moessner-13}.
The Bloch Hamiltonian matrix is given by
\begin{align}
    H(\bm{k})=
    \begin{pmatrix}
    2t_2(\cos k_1-\cos k_2)&t_1f^*(\bm{k})\\
    t_1f(\bm{k})&-2t_2(\cos k_1-\cos k_2)\\
    \end{pmatrix},
\end{align}
where
\begin{align}
    &f(\bm{k})=e^{-i\pi/4}\left[1+e^{i(k_2-k_1)}\right]+e^{i\pi/4}\left[e^{-ik_1}+e^{ik_2}\right].
\end{align}
The unit cell is a square lattice and consists of two local orbitals, labeled 1 and 2.
As a starting point, we consider the parameter set $(t_1,t_2)=(1,1/\sqrt{2})$, for which the lowest-energy band becomes maximally flat and carries a unit Chern number \cite{Neupert-Santos-Chamon-Mudry-11}.
In the following, we consider the projection onto the lowest Chern band.
As an interaction, we first consider a nearest-neighbor repulsive interaction between the local orbitals 1 and 2.
The interaction strength is set to unity.
Under this setup, the independent two-particle orbitals are given by the following sets [see also 
Fig. \ref{fig2}(a)]:
\begin{align}
    \{i,j,\Delta\bm{r}\}=&\{1,2,(0,0)\},\{1,2,(-1,0)\},\{1,2,(0,1)\},\notag\\
    &\{1,2,(-1,1)\}.
\end{align}
Thus, the small Hamiltonian $\hat{h}_{\bm{q}}$ is a $4\times4$ matrix.
In the two-particle problem, there are four bound states for each total momentum $\bm{q}$.
The bound-state energies are obtained by diagonalizing the small Hamiltonian [Fig. \ref{fig2}(b)].
The system size used in this calculation is $16\times16$.
The two-particle band structure splits into two high-energy bands and two low-energy bands.

Let us consider the two-particle band topology of the upper bands. These two bands are degenerate along the Brillouin zone boundary and cannot be separated.
Therefore, we compute the Berry curvature of the two bands collectively [Fig. \ref{fig2}(c)].
For the calculation, we have used the Fukui-Hatsugai-Suzuki formula \cite{Fukui-Hatsugai-Suzuki-05}.
By integrating the Berry curvature, we find that the two dominant two-particle bands carry a unit Chern number.
Since the total Chern number of all two-particle bands must be zero, the remaining near-zero-energy bands must carry a total Chern number of $-1$.

Next, we investigate the relationship between the two-particle band structure and the many-body ground states at filling $\nu=1/3$.
We perform exact diagonalization for a system of $6 \times 4$ unit cells.
At this filling, the number of particles, $N_{\rm p}$, is $8$.
This model has been studied as a candidate for an FCI \cite{Neupert-Santos-Chamon-Mudry-11}.
Indeed, our numerical calculations also show a threefold ground-state degeneracy [Fig. \ref{fig3}].
In general, it is nontrivial to determine whether such a degeneracy originates from topological order. However, since this model is an established candidate for an FCI, we proceed under the assumption that the observed threefold degeneracy reflects the topological ground-state degeneracy characteristic of topologically ordered phases \cite{Wen-Niu-90}.
At this parameter point, the nontrivial two-particle topology of the two dominant bands and the FCI ground states at $\nu=1/3$ are simultaneously realized.
We now turn to the case where the hopping parameters and the interaction terms are varied.

We first change the ratio $r_t=\sqrt{2}t_2/t_1$.
The case with the flattest single-particle band discussed above corresponds to $r_t=1$.
Although the single-particle band dispersion does not directly affect the physics after projection, $r_t$ still affects the topological and geometrical properties of the one-particle states.
As $r_t$ is increased, the upper two and lower two bands of the bound states in the two-particle spectrum gradually hybridize [Fig. \ref{fig3}(a)].
In the many-body spectrum, the gap between the threefold degenerate ground states and the excited states gradually decreases [Fig. \ref{fig3}(a)].
At $r_t=3$, the threefold degeneracy merges into the continuum of excited states. 
These results suggest a correlation between the isolation of the two dominant bands in the two-particle spectrum and the presence of the threefold ground-state degeneracy in the many-body spectrum.

Next, we change the shape of the interaction.
Thus far, we have considered the case where each site interacts equally with its four nearest neighbors [Fig. \ref{fig2}(a)]. 
Among the four interactions, one occurs within the unit cell, while the remaining three occur between different unit cells.
We denote the strength of the former interaction by $U_1$, and that of the latter by $U_2$.
Fixing $U_1=1$, we vary the ratio $r_U=U_2/U_1$.
For $U_2=0$, the interaction occurs only within the unit cell, and the number of the relevant two-particle orbitals is reduced from four to one:
\begin{align}
    \{i,j,\Delta\bm{r}\}=&\{1,2,(0,0)\}.
\end{align}
Thus, the small Hamiltonian is a $1\times1$ matrix, implying that the bound-state band cannot have a topological structure.
As $r_U$ is decreased from 1 to 0, one of the two upper bands gradually separates in energy and is eventually absorbed into the zero-energy manifold [Fig. \ref{fig3}(b)]. As predicted by the above analysis, this isolated band carries a unit Chern number.
During this parameter variation, the threefold degeneracy becomes separated from the other states. However, the excitation gap rapidly decreases and is reduced to $\mathcal{O}(10^{-3})$ at $U_2=0$.
A similar result involving tiny-gap ground states has been observed in the QWZ model with the same interaction \cite{Wu-Bernevig-Regnault-12}.
In our formalism, the corresponding small Hamiltonian for this model is also a $1\times1$ matrix, as in our calculation with $r_U=0$.
The above results strongly suggest that the presence of the two-particle topological band, or a topological interaction, is closely related to the magnitude of the excitation gap.

\section{Two-body topological obstruction to trivial ground state \label{sec: obstruction}}
\begin{figure*}[]
\begin{center}
 \includegraphics[width=14cm,angle=0,clip]{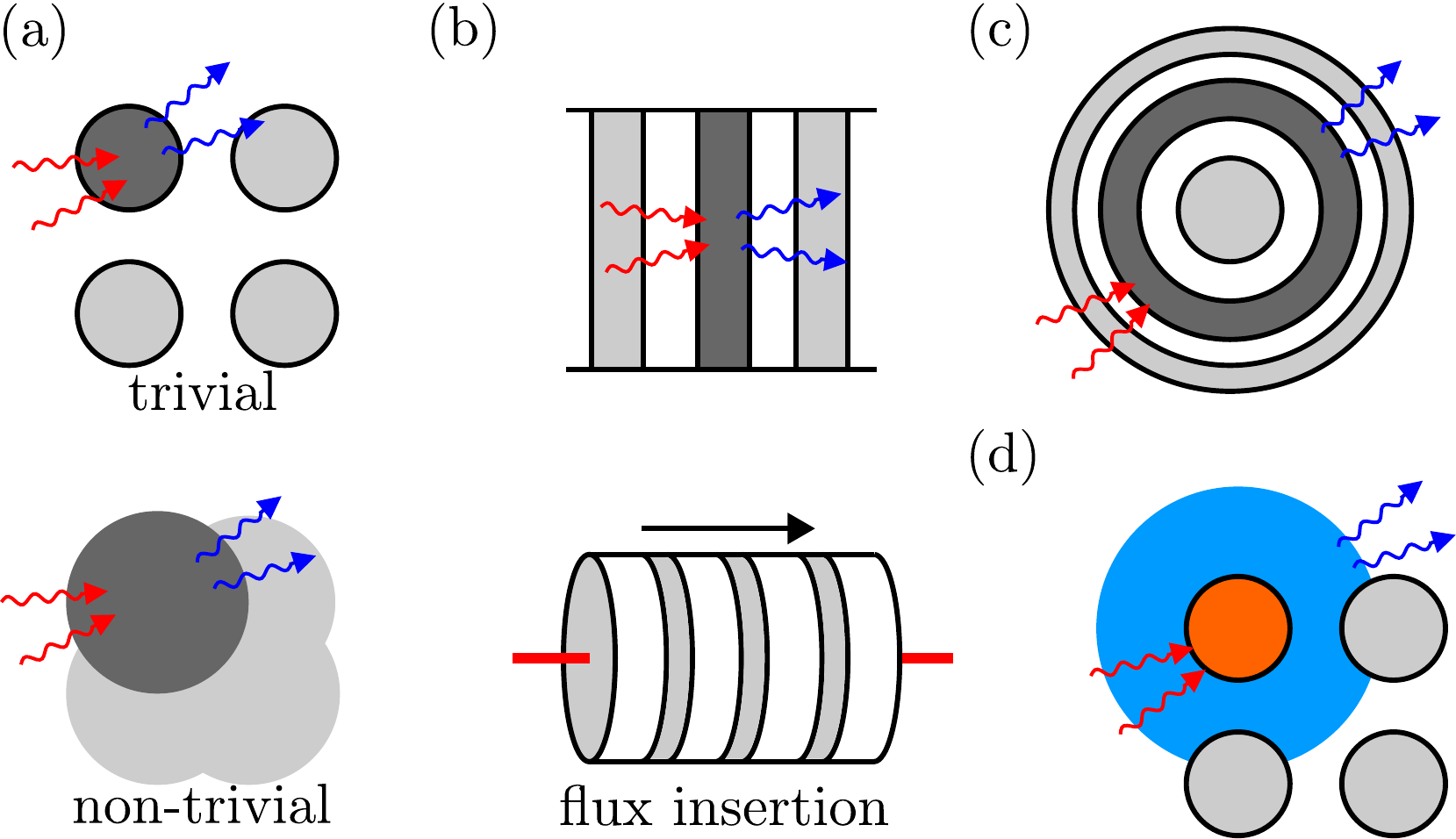}
 \caption{Schematic pictures of scattering basis in effective interactions. Red and blue arrows represent incoming and outgoing particles, respectively. (a) Wannier basis. Under the topological obstruction, the scattering channels become widely distributed. (b) Cylindrical basis. On this basis, the scattering can be interpreted as occurring in a one-dimensional manner. Under the flux insertion, the interaction itself is topologically pumped. (c) Radially-localized basis. This basis provides an analogy to the analytical structure found in Landau levels. (d) Biorthogonal basis. On this basis, two particles scatter into localized channels and are emitted from delocalized channels. } 
 \label{fig4}
\end{center}
\end{figure*}

As discussed in Sec. \ref{sect:pseudo}, the two-particle topology of the small Hamiltonian $\hat{h}_{\bm{q}}$ is an important factor for realizing the FCI, although it is not sufficient on its own to ensure their emergence. 
Before concluding this paper, we take a different perspective and discuss how the topological two-body interaction can serve as an obstruction to trivial ground states.

We begin with the scattering elements:
\begin{align}
    [\hat{S}^{\dagger}_{\bm{q}}\hat{S}_{\bm{q}}]_{\bm{k}_1,\bm{k}_2}=\bra{\bm{q};\bm{k}_1}
    \left[\sum_{\bm{q}',\beta}\ket{\Psi_{\bm{q}',\beta}}\epsilon_{\bm{q}',\beta}\bra{\Psi_{\bm{q}',\beta}}\right]
    \ket{\bm{q};\bm{k}_2}.
\end{align}
Suppose that the dominant two-particle energies are equal to 1, while the remaining two-particle energies are negligibly small.
Then, the scattering elements are approximately given in terms of the two-particle projection matrix:
\begin{align}
    [\hat{S}^{\dagger}_{\bm{q}}\hat{S}_{\bm{q}}]_{\bm{k}_1,\bm{k}_2}&=\bra{\bm{q};\bm{k}_1}\hat{P}^{(2p)}\ket{\bm{q};\bm{k}_2},\label{effective}\\
    \hat{P}^{(2p)}&=\sum_{\bm{q}',n}\ket{\Psi_{\bm{q}',n}}\bra{\Psi_{\bm{q}',n}},
\end{align}
where the summation is taken over the highest-energy bands indexed by $n$.
While the original repulsive interaction takes place between two nearby points, $\{\bm{r},i\}$ and $\{\bm{r}-\Delta\bm{r},j\}$, the effective interaction (\ref{effective}) can be considered to take place in the two-particle effective scattering channels defined by the eigenbasis of the projection operator $\hat{P}^{(2p)}$.

There are many possible choices of such a basis.
Here, we focus on the spatial locality of the interaction and adopt the ``two-particle Wannier functions" as the basis functions:
\begin{align}
    \hat{P}^{(2p)}=\sum_{\bm{R},n}\ket{w_{n,\bm{R}}}\bra{w_{n,\bm{R}}},
\end{align}
where $\bm{R}$ denotes the unit-cell vector.
If the dominant two-particle bands are topologically trivial, all $\ket{w_{n,\bm{R}}}$'s are exponentially localized around $\bm{R}$. 
This indicates that the projection onto the band of interest, $\mathcal{P}_{\alpha}$, does not significantly affect the interaction range, and the resulting effective interaction remains simple and short-ranged.
In the case of sufficiently large filling, the ground state is expected to favor metals or gapped states that break translational symmetry, such as charge-ordered states or Wigner crystal-like states.

However, if the dominant two-particle bands carry a finite Chern number, the situation changes qualitatively.
It is well established that an exponentially-localized Wannier basis cannot be constructed for a Chern band \cite{thouless1984wannier,rashba1997orthogonal,thonhauser2006insulator}. In this case, the two-particle Wannier function $\ket{w_{n,\bm{R}}}$ exhibits only algebraic localization.
This means that the effective interaction cannot be reduced to a simple short-range form [Fig. \ref{fig4}(a)].
Accordingly, simple symmetry-broken states cannot account for the ground state of the effective Hamiltonian.

In the following, we explore qualitative pictures by considering various bases of the two-particle Chern bands.
Note that in order to make any definitive statements about the ground state, it is essential to understand the detailed matrix elements between $\{\ket{\bm{q};\bm{k}}\}$ and the scattering basis.

\subsection{Cylindrical basis}
While an exponentially localized Wannier basis in both the $x$ and $y$ directions cannot be constructed, it is possible to construct a one-dimensional Wannier basis (cylindrical basis) that is exponentially localized in either the $x$ or $y$ direction (see review paper \cite{Bergholtz-Liu-13} for summaries of several relevant works).
In this case, the effective interaction exhibits a one-dimensional character [Fig. \ref{fig4}(b)].
Moreover, by inserting flux through the handle of the torus, the effective interaction itself undergoes a topological pump [Fig. \ref{fig4}(b)].
If the one-particle band also possesses a nonzero Chern number, it is possible to construct a one-dimensional basis in which the one-particle states also undergo a topological pump.
Since all relevant ingredients undergo the topological pump, the ground state may be strongly influenced by it.
This sensitivity to flux insertion is a hallmark of FCIs \cite{Bergholtz-Liu-13}.
In summary, the coexistence of one-particle and two-particle topologies provides strong evidence characterizing FCIs.

\subsection{Radially-localized basis}
In the physics of the FQHE, the analytic property with respect to the complex coordinate $z=x+iy$ plays a crucial role in dramatically simplifying the many-body nature of the problem.
Recently, the notion of a vortex function $z(x,y)$ has been introduced to describe the distance between the Chern band of interest and the Landau levels \cite{ledwith2023vortexability,fujimoto2025higher}.
The vortex function can be a nonlinear function of the real-space coordinates and is capable of describing complex phenomena in solid-state physics. One of the authors has generalized this concept to lattice systems using tight-binding expressions \cite{okuma2024constructing,okuma2025biorthogonal}.
In this formalism, all nonlinearity is absorbed into the virtual sublattice positions $\tilde{\bm{r}}_i=(\tilde{x}_i,\tilde{y}_i)$, and the lattice vortex function can then be expressed as a linear function of the virtual position operator:
\begin{align}
    Z&=\alpha_1~ (X+\sum_i\tilde{x}_iP_i)+\alpha_2~ (Y+\sum_i\tilde{y}_iP_i)\notag\\
    &=\alpha_1~ X+\alpha_2~ Y +\sum_i\gamma_iP_i,
\end{align}
where $\bm{R}=(X,Y)$ is a unit-cell position operator, $\alpha_1,\alpha_2,\gamma_i$ are complex parameters, and $P_i:=\ket{i}\bra{i}$ is the projection operator onto sublattice $i$. 
The difference between the actual and virtual sublattice positions captures the nonlinearity mentioned above.
The parameters of the lattice vortex function, $\alpha_1,\alpha_2,\gamma_i\in\mathbb{C}$ should be determined based on some criteria, such as ideal conditions \cite{parameswaran2012fractional,Roy-geometry-14,Jackson-Moller-Roy-15,Claassen-Lee-Thomale-Qi-Devereaux-15, Lee-Claassen-Thomale-17} and vortexability \cite{ledwith2023vortexability,fujimoto2024higher,okuma2024constructing}.
Using the lattice vortex function, the radially-localized basis is defined by the eigenstates of the operator: 
\begin{align}
    &PZPZ^*P\ket{\phi_m}=\lambda_m\ket{\phi_m},\notag\\
    &\ket{\phi_m}=\sum_{\bm{k}}\frac{1}{\sqrt{N_{\rm unit}}}a_{m}(\bm{k})\ket{\bm{k},\alpha},
\end{align}
where $\{\ket{\bm{k},\alpha}\}$ is the set of the Bloch eigenstates, and $P=\sum_{\bm{k}}\ket{\bm{k},\alpha}\bra{\bm{k},\alpha}$ is the projection operator onto the Chern band $\alpha$.
For more details, see Refs. \cite{okuma2024constructing,okuma2025biorthogonal}.
The corresponding operator in the Landau level is twice the angular momentum operator, and the eigenstates form the analytic basis $\{\phi_m(z)\propto z^m\exp(-|z|^2/4)\}$.
A similar basis, known as the momentum-space Landau level, was constructed in Refs. \cite{Claassen-Lee-Thomale-Qi-Devereaux-15,Lee-Claassen-Thomale-17}.
In the context of the lattice vortex function, it corresponds to the eigenbasis of the operator $PZZ^*P$.

The concept of the vortex function can be naturally extended to the two-particle Chern band.
A two-particle orbital is represented by the coordinate $(\bm{r}, i)$, where one particle is located at $(\bm{r}, i)$ and the other at $(\bm{r} - \Delta \bm{r}, j)$.
For simplicity, we set $(\alpha_1,\alpha_2)=(1,i)$ and neglect the nonlinearlity.
We then define the two-particle radially localized basis [Fig. \ref{fig4}(c)] as follows:
\begin{align}
    &\hat{P}^{(2p)}Z^{(2p)}\hat{P}^{(2p)}Z^{*(2p)}\hat{P}^{(2p)}\ket{M}=\lambda_M\ket{M},\\
    &Z^{(2p)}=(x+\sum_i x_iP_i)+i~ (y+\sum_iy_iP_i),\\
    &\ket{M}=\sum_{\bm{q},n}\frac{1}{\sqrt{N_{\rm unit}}}a^{(2p)}_{M}(\bm{q},n)\ket{\Psi_{\bm{q},n}}.
\end{align}
where $i$ is the first component of the two-particle orbital, $a=\{i,j,\Delta\bm{r}\}$.
The index $M$ is a direct analogy of the center-of-mass angular momentum.
Note that the information about the relative angular momentum is encoded in the two-particle Chern band index $n$. 
The real-space representations of these states are given by
\begin{align}
    \Psi_{M}(\bm{r},\{i,j,\Delta\bm{r}\})=\sum_{\bm{q},n}\frac{a^{(2p)}_{M}(\bm{q},n)}{N_{\rm unit}}e^{i\bm{q}\cdot\bm{r}}\psi_{\bm{q},n}(\{i,j,\Delta\bm{r}\}).
\end{align}
This is the direct analogy of the two-particle wave function in the Landau level problem, $\Psi_{M,m}(Z,z)$ in Eq. (\ref{two-particle-landau}).

In the presence of both one- and two-particle topologies, all scattering information can be expressed using a radially localized basis that mimics the angular-momentum eigenstates in the Landau levels.
This fact is expected to strongly support the realization of FCIs.

\subsection{Biorthogonal basis}
The two bases discussed above are tailored to the typical geometries of the FQHE: the cylinder and the disc.
In the following, we consider the scenario in which two particles scatter into spatially localized and periodically structured channels.
Our focus here is not on verifying the existence of FCIs, but rather on elucidating the unusual nature of the topological interaction.

Instead of the non-local Wannier functions, we consider the coherent-like states introduced in Refs. \cite{okuma2024constructing,okuma2025biorthogonal}.
We first consider the case of one-particle topology, but the two-particle case can be discussed in a similar manner.
The coherent-like states are defined as the eigenstates of $PZ^*P$, and can be constructed from $\ket{\phi_{m=0}}$, which is shown to be the topologically protected zero mode of $PZ^*P$ \cite{okuma2024constructing,okuma2025biorthogonal}:
\begin{align}
&PZ^*P\ket{\zeta_{\bm{R}}}=[Z_{\bm{R}}]^*\ket{\zeta_{\bm{R}}},\\
    &\ket{\zeta_{\bm{R}}}=\sum_{\bm{k}}\frac{e^{-i\bm{k}\cdot\bm{R}}}{\sqrt{N_{\rm unit}}}a_{0}(\bm{k})\ket{\bm{k}},
\end{align}
where $Z_{\bm{R}}=\alpha_1X+\alpha_2 Y$, and the band index $\alpha$ is omitted. The set $\{Z_{\bm{R}}\}$ forms a lattice structure in the complex plane, which is an analogy of the von Neumann lattice \cite{von2018mathematical,perelomov2002completeness, bargmann1971completeness,imai1990field,ishikawa1992field,ishikawa1999field}.
At a certain momentum $\bm{k}$, $a(\bm{k})$ vanishes. We assume $a_0(\bm{k}=0)=0$. 
The resulting coherent-like state is exponentially localized, at the cost of orthogonality.
We then define the conjugate basis $\{\ket{Z_{\bm{R}}}\}$ as
\begin{align}
    \ket{Z_{\bm{R}}}=\sum_{\bm{k}\neq\bm{0}}\frac{e^{-i\bm{k}\cdot\bm{R}}}{\sqrt{N_{\rm unit}}}\frac{1}{[a_{0}(\bm{k})]^*}\ket{\bm{k}},
\end{align}
which satisfies $\bra{Z_{\bm{R}}}\zeta_{\bm{R}'}\rangle=\delta_{\bm{R},\bm{R}'}-1/N_{\rm unit}$.
In the infinite-volume limit, the sets $\{\ket{\zeta_{\bm{R}}}\}$ and $\{\ket{Z_{\bm{R}}}\}$ form a biorthogonal basis \cite{okuma2024constructing,okuma2025biorthogonal}.
In a finite system, the projection operator is expressed as
\begin{align}
    P&=\ket{Z_{\bm{R}}}\bra{\zeta_{\bm{R}}}+\frac{\ket{\bm{k}=0}}{\sqrt{N_{\rm unit}}}\bra{\zeta_{\bm{R}}}\notag\\
    &+\ket{Z_{\bm{R}}}\frac{\bra{\bm{k}=0}}{\sqrt{N_{\rm unit}}}+\frac{\ket{\bm{k}=0}\bra{\bm{k}=0}}{N_{\rm unit}}.
\end{align}
Here we have used the complete biorthogonality of $\{\ket{\zeta_{\bm{R}}}+\ket{\bm{k}=0}/\sqrt{N_{\rm unit}}\}$ and $\{\ket{Z_{\bm{R}}}+\ket{\bm{k}=0}/\sqrt{N_{\rm unit}}\}$.
Note that the conjugate basis $\{\ket{Z_{\bm{R}}}\}$ is algebraically localized. 

The same construction can be extended to the two-particle Chern band [Fig. \ref{fig4}(d)]:
\begin{align}
    \hat{P}^{(2p)}&=\ket{Z^{(2p)}_{\bm{R}}}\bra{\zeta^{(2p)}_{\bm{R}}}+\frac{\ket{\bm{q}=0}}{\sqrt{N_{\rm unit}}}\bra{\zeta^{(2p)}_{\bm{R}}}\notag\\
    &+\ket{Z^{(2p)}_{\bm{R}}}\frac{\bra{\bm{q}=0}}{\sqrt{N_{\rm unit}}}+\frac{\ket{\bm{q}=0}\bra{\bm{q}=0}}{N_{\rm unit}}.
\end{align}
Here, we omit the contribution from the trivial two-particle band.
If two particles scatter into the exponentially localized channels $\{\ket{\zeta^{(2p)}_{\bm{R}}}\}$, they are emitted from the conjugate channels $\{\ket{Z^{(2p)}_{\bm{R}}}\}$ or $\ket{\bm{q}=0}$, both of which are not exponentially localized.
This nonreciprocal nature of the scattering is a distinctive feature of the topological interaction.

\section{Conclusions\label{sec:conclusion}}
In this paper, we investigated the role of the topology of two-body interactions in the realization of FCIs.
We defined the matrix $\hat{S}_{\bm{q}}$, which characterizes the effective interaction projected onto the band, and formulated the two-particle problem using $\hat{S}_{\bm{q}}$.
This approach can be applied to both fermions and bosons by switching the symmetry of $\hat{S}_{\bm{q}}$.
We also focused on the pseudopotential in the FQHE problem and qualitatively discussed that the small Hamiltonian $h_{\bm{q}}=\hat{S}_{\bm{q}}\hat{S}^{\dagger}_{\bm{q}}$, whose eigenvalues are the energy of the bound states, should possess two-particle Chern bands.
The relationships between the exact diagonalization spectra at $\nu=1/3$ filling and the two-particle band structures support this claim.
Finally, by decomposing the topological two-body interaction into scattering channels, we proposed an intuitive picture in which the interaction itself hosts the Wannier obstruction and thereby acts to prevent a trivial ground state.

The notion of topological two-body interaction introduced in this paper opens up several potential avenues for future research.
\begin{itemize}
    \item By investigating how the interplay between single-particle topological and geometrical structures and interactions influences the two-body topological structure, one may identify favorable conditions for realizing FCIs. 
    Focusing on the relationship between two-particle bands and two-particle Landau levels, it is interesting to consider the properties of the quantum geometric tensor defined for the two-particle bands and their connection to the stability of FCIs. Similar to the case of single-particle bands, the trace condition may play an important role.
    Whether FCIs with higher Chern numbers or non-Abelian FCIs can be characterized by topological two-body interactions remains a nontrivial open question for future work.
    \item It is of interest to explore what types of phases may emerge from topological two-body interactions in systems without single-particle topology or at arbitrary filling factors.
    At sufficient filling, electrons are expected to occupy near-zero-energy topological (delocalized) bound states, which may lead to exotic states such as superconductivity. 
    \item It is also of interest to examine what phases may emerge when the two-particle topology is characterized by a $\mathbb{Z}_2$ invariant, and to explore potential connections to fractional topological insulators \cite{levin2009fractional,stern2016fractional}.
\end{itemize}
We believe that these issues represent intriguing directions in the exploration of strongly correlated phenomena in flat-band systems.
\acknowledgements
The authors thank Koji Kudo for the insightful discussions that contributed to a deeper understanding of the projection Hamiltonian.
This work was supported by JSPS KAKENHI Grant No.~JP20K14373 and No.~JP23K03243.
\\
\appendix
\section{Formalism for bosons}
There are mainly two differences in the formulation for bosons as opposed to fermions. The first one is the difference between symmetry and antisymmetry of $\hat{S}_{\bm{q}}$.
In the bosonic case, we can symmetrize the Hamiltonian matrix elements:
\begin{align}
    \mathcal{P}_{\alpha}H_{\rm int}\mathcal{P}_{\alpha}&=\frac{1}{2}\sum_{\bm{k}_1,\bm{k}_2,\bm{q}}[\hat{S}^{\dagger}_{\bm{q}}\hat{S}_{\bm{q}}]_{\bm{k}_1,\bm{k}_2}c^{\dagger}_{\bm{k}_1}c^{\dagger}_{\bm{q}-\bm{k}_1}c_{\bm{q}-\bm{k}_2}c_{\bm{k}_2},\\
    [\hat{S}_{\bm{q}}]_{a,\bm{k}}&=\frac{[\hat{A}_{\bm{q}}]_{a,\bm{k}}+[\hat{A}_{\bm{q}}]_{a,\bm{q}-\bm{k}}}{\sqrt{2}}.
\end{align}
The bosonic symmetry is encoded into the following relation:
\begin{align}
    [\hat{S}_{\bm{q}}]_{a,\bm{k}}=[\hat{S}_{\bm{q}}]_{a,\bm{q}-\bm{k}}~.\label{bosonsymmetry}
\end{align}

Another important difference is the absence of the Pauli principle. 
In the bosonic case, $\bm{k}\equiv\bm{q}-\bm{k}$ modulo a reciprocal lattice vector is allowed. 
Thus, the following basis spans the two-particle Hilbert space:
\begin{align}
    \ket{\bm{q};\bm{k}}:=
    \begin{cases}
    c^{\dagger}_{\bm{q}-\bm{k}}c^{\dagger}_{\bm{k}}\ket{0}&\mathrm{for}~n(\bm{k})< n(\bm{q}-\bm{k})\\
    c^{\dagger}_{\bm{k}}c^{\dagger}_{\bm{k}}\ket{0}/\sqrt{2}&\mathrm{for}~n(\bm{k})= n(\bm{q}-\bm{k})
    \end{cases}~~~~.
\end{align}
For this reason, the large Hamiltonian is modified as follows.
\begin{align}
&[\hat{\mathcal{H}}_{\bm{q}}]_{\bm{k},{\bm{k}'}}:=\bra{\bm{q};\bm{k}}\mathcal{P}_{\alpha}H_{\rm int}\mathcal{P}_{\alpha}\ket{\bm{q};\bm{k}'}=2~[\hat{S}'^{\dagger}_{\bm{q}}\hat{S}'_{\bm{q}}]_{\bm{k},\bm{k}'},\notag\\
&[\hat{S}'_{\bm{q}}]_{a,\bm{k}}=
    \begin{cases}
[\hat{S}_{\bm{q}}]_{a,\bm{k}}&\mathrm{for}~n(\bm{k})< n(\bm{q}-\bm{k})\\
    [\hat{S}_{\bm{q}}]_{a,\bm{k}}/\sqrt{2}&\mathrm{for}~n(\bm{k})= n(\bm{q}-\bm{k})
    \end{cases}~~~~.
\end{align}
While the large Hamiltonian is modified, the small Hamiltonian has the same expression as in the case of fermions:
\begin{align}
    &[\hat{h}_{\bm{q}}]_{a,a'}:=\sum_{\{\bm{k}|n(\bm{k})\leq n(\bm{q}-\bm{k})\}}2[\hat{S}'_{\bm{q}}]_{a,\bm{k}}[\hat{S}'^{\dagger}_{\bm{q}}]_{\bm{k},a'}\notag\\
    =&\sum_{\{\bm{k}|n(\bm{k})<n(\bm{q}-\bm{k})\}}[\hat{S}_{\bm{q}}]_{a,\bm{k}}[\hat{S}^{\dagger}_{\bm{q}}]_{\bm{k},a'}
    +[\hat{S}_{\bm{q}}]_{a,\bm{q}-\bm{k}}[\hat{S}^{\dagger}_{\bm{q}}]_{\bm{q}-\bm{k},a'}\notag\\
    +&\sum_{\{\bm{k}|n(\bm{k})=n(\bm{q}-\bm{k})\}}2\times\left(\frac{1}{\sqrt{2}}\right)^2[\hat{S}_{\bm{q}}]_{a,\bm{k}}[\hat{S}^{\dagger}_{\bm{q}}]_{\bm{k},a'}\notag\\
    =&\sum_{\bm{k}}[\hat{S}_{\bm{q}}]_{a,\bm{k}}[\hat{S}^{\dagger}_{\bm{q}}]_{\bm{k},a'}.
\end{align}

\section{Formalism for multiple target bands}
The extension for the case where there are multiple target bands for the projection is almost trivial.
Let us consider the following projection Hamiltonian:
\begin{align}
    \mathcal{P}H_{\rm int}\mathcal{P}
    =&\frac{1}{2N_{\rm unit}}\sum_{i,j}\sum_{\bm{q},\bm{k}_1,\bm{k}_2,\Delta\bm{r}}e^{-i(\bm{k}_1-\bm{k}_2)\cdot\Delta\bm{r}}V^{ij}(\Delta\bm{r})\notag\\
    &\sum_{\alpha,\beta,\gamma,\delta}u^{*}_{\bm{k}_1,\alpha}(i)u^{*}_{\bm{q}-\bm{k}_1,\beta}(j)u_{\bm{q}-\bm{k_2},\gamma}(j)u_{\bm{k}_2,\delta}(i)\notag\\
    &c^{\dagger}_{\bm{k}_1,\alpha}c^{\dagger}_{\bm{q}-\bm{k}_1,\beta}c_{\bm{q}-\bm{k}_2,\gamma}c_{\bm{k}_2,\delta},
\end{align}
where $\mathcal{P}$ is the projection operator onto the bands of interest.
The following basis spans the two-particle Hilbert space:
\begin{align}
    \ket{\bm{q};\bm{k},\alpha\beta}:=c^{\dagger}_{\bm{q}-\bm{k},\beta}c^{\dagger}_{\bm{k},\alpha}\ket{0}~~\mathrm{for}~n(\bm{k},\alpha)< n(\bm{q}-\bm{k},\beta),
\end{align}
where $n(\cdot)$ labels the set $\{\bm{k},\alpha\}$.
The matrix $\hat{A}_{\bm{q}}$ and $\hat{S}_{\bm{q}}$ are extended as follows:
\begin{align}
    &[\hat{A}_{\bm{q}}]_{a,\bm{k}\alpha\beta}=\sqrt{\frac{V^{ij}(\Delta\bm{r})}{N_{\rm unit}}}e^{i\bm{k}\cdot\Delta\bm{r}}u_{\bm{k},\alpha}(i)u_{\bm{q}-\bm{k},\beta}(j),\notag\\
    &[\hat{S}_{\bm{q}}]_{a,\bm{k}\alpha\beta}=\frac{[\hat{A}_{\bm{q}}]_{a,\bm{k}\alpha\beta}-[\hat{A}_{\bm{q}}]_{a,\bm{q}-\bm{k}\beta\alpha}}{\sqrt{2}}.
\end{align}
With this new $\hat{S}_{\bm{q}}$, one can define the scattering matrix $\hat{S}^{\dagger}_{\bm{q}}\hat{S}_{\bm{q}}$ and the small/large Hamiltonians, following the formulation in the main text.

\bibliography{FCI}
\end{document}